\documentclass[aps,prb,amsmath,amssymb,twocolumn,groupedaddress]{revtex4-2}
\usepackage[]{graphicx}
\usepackage{dcolumn}
\usepackage{multirow}
\usepackage{bm}

\begin{document}

\title{Metallic and semimetallic states of molecular crystalline hydrogen at high pressures.}

\author{I. Saitov}
\email[]{E-mail: saitovilnur@gmail.com}
\affiliation{Joint Institute for High Temperatures of the Russian Academy of Sciences (JIHT RAS), Izhorskaya 13 Building 2, Moscow 125412, Russia}
\affiliation{Moscow Institute of Physics and Technology National Research University (MIPT NRU), Institutskij pereulok 9, Dolgoprudny, Moscow Region 141700, Russia}
\affiliation{National Research University Higher School of Economics (NRU HSE), Myasnitskaya Ulitsa 20, Moscow 101000, Russia}

\date{\today}

\begin{abstract}
Ab initio molecular dynamic method within the framework of density functional theory is applied to analyze the structural and electronic properties of crystalline molecular hydrogen at temperature 100\,K. Pressure, pair correlation function and band structure are calculated. The crossover of molecular crystalline hydrogen from the state of a semiconductor to a semimetallic and metallic state is observed upon compression in the pressure range of 302-626\,GPa. At pressures below 361\,GPa, the molecular crystal with the C2/c structure is a semiconductor with an indirect gap. In the pressure range 361 - 527\,GPa, band structure of the monoclinic C2/c lattice has a characteristic semimetalic profile with partially unoccupied valence band and partially occupied conduction band. When compressed to pressures above 544\,GPa, the structure changes from monoclinic C2/c to orthorhombic Cmca, accompanied by a sharp decrease (by more than two orders of magnitude) in the value of the direct gap, which is an indication of the metallic conductivity of the resulting structure. The metallic state is metastable and exists up to a pressure of 626\,GPa.
\end{abstract}

\maketitle

\section{Introduction}
The metallization of crystalline hydrogen at high pressures has been theoretically investigated since 1935. In the original work \cite{WignerHuntington1935}, it is assumed that hydrogen molecules dissociate, forming a body-centered cubic lattice with one proton per unit cell. The critical temperature of the transition of solid hydrogen to the superconducting state is estimated at about 400\,K \cite{Ashcroft1968}, which is a clear indication of the high-temperature superconductivity of crystalline hydrogen at pressures up to normal, due to the wide range of the metastable states \cite{Kagan1972,Brovman1972,Brovman1972a,Brovman1973,Kagan1977}.

There are two suggested mechanisms of the formation of metallic solid hydrogen. The first mechanism is based on the prediction of \cite{WignerHuntington1935} and is associated with the dissociation of hydrogen molecules with the formation of an atomic lattice. This mechanism is studied theoretically within the framework of various \textit{ab initio} approaches based on the density functional theory (DFT) \cite{Pickard2007,McMahon2011,Degtyarenko2016,Kudryashov2016,Degtyarenko2017,Kudryashov2017,Zhang2018,Rillo2018} and the quantum Monte-Carlo method \cite{Azadi2014,McMinis2015}, which predict the formation of the atomic metallic crystalline hydrogen in the pressure range of $P=370–-500$\,GPa. Experimentally, the formation of crystalline metallic hydrogen is presumably observed in \cite{Dias2017}, where the reflectivity of solid hydrogen is studied under static compression in the diamond anvil cell at temperatures of 5.5 and 83\,K. A sharp increase of the reflectivity is detected at the pressure 495\,GPa, which is interpreted as a transition to the metallic state. In addition to \cite{Dias2017}, an indication of the metallization of the solid hydrogen is also found in \cite{Eremets2019,Loubeyre2020} at pressures of 440 and 425\,GPa, respectively. The analysis of the results of \cite{Dias2017} using the Drude formula for the calculation of the dielectric function provides the estimation of the degree of ionization, which reaches a value close to 100\% at the transition pressure. The result indicates that the obtained metallic phase is an atomic one.

The second mechanism assumes that the metallization of the crystalline hydrogen is associated with overlapping of the conduction and valence bands and structural transformations, but the hydrogen crystal remains molecular. The bandgap closure, as well as the smooth growth of electrical conductivity in the phase III of molecular crystalline hydrogen, are obtained theoretically within the framework of the DFT and the quantum Monte-Carlo approaches in \cite{Pickard2007,Johnson2000,Cudazzo2008,Lebegue2012,Pickard2012,Azadi2013,Goncharov2013,Drummond2015,Monserrat2016,Azadi2017,Azadi2018,Norman2019,Saitov2019,Saitov2020,Norman2020}. According to the results of \cite{Pickard2007}, the most probable structure of crystalline molecular hydrogen at pressures above 200\,GPa in the phase III is a monoclinic lattice with the C2/c symmetry with 12 atoms per unit cell. According to the estimates \cite{Johnson2000,Cudazzo2008,Lebegue2012,Pickard2012,Azadi2013,Goncharov2013,Drummond2015,Monserrat2016,Azadi2017,Azadi2018}, metallization of the molecular structure occurs in the range of $P=250–500$\,GPa. A monoclinic lattice with C2/c symmetry and two orthorhombic lattices with Cmca-12 and Cmca-4 symmetries \cite{Zhang2018,Rillo2018,Azadi2014,McMinis2015,Cudazzo2008,Lebegue2012,Pickard2012,Azadi2013,Goncharov2013,Drummond2015,Monserrat2016,Azadi2017,Azadi2018} containing 12 and 4 hydrogen atoms per unit cell are considered as the most probable structures of the metallic molecular crystalline hydrogen.

In the experiments \cite{Eremets2019,Loubeyre2020}, in contrast to \cite{Dias2017}, it is suggested that the resulting metallic state of crystalline hydrogen retains its molecular structure. In work \cite{Eremets2019}, the temperature dependence of the resistivity and the profiles of the Raman spectrum are studied in the range of parameters $T<200$\,K and $P=195 - 480$\,GPa. The obtained results indicate to the existence of a semimetallic state of molecular crystalline hydrogen at $P=350 - 440$\,GPa. In the given pressure range, the temperature dependence of resistivity has a minimum, which is characteristic for semimetals. The measured absolute values of conductivity are also typical for semimetals, in particular, for bismuth \cite{Dorofeev1984,Brown2015}. The profile of the pressure dependence of the resistivity of molecular hydrogen is similar to the analogous dependence for xenon and oxygen \cite{Shimizu2018,Eremets2000,Koufos2015}, in which the metallization also passes through the formation of an intermediate semimetallic state. It should be noted that no structural changes are observed in the range $P=350 - 440$\,GPa and crystalline hydrogen remains in the phase III, with the C2/c structure. In the Raman spectrum, the peak corresponding to the hydrogen molecule disappears at pressures above 440\,GPa. Authors \cite{Eremets2019} suggest that this can be considered as an indication of further transformation of solid molecular hydrogen both into a good molecular metal and into an atomic state.

A sharp decrease of the direct gap from 0.6 to 0.1\,eV is observed at pressure of 425\,GPa and temperature of 80\,K in the experiment \cite{Loubeyre2020}. This result is considered as an indication of the formation of the metallic state. The results \cite{Loubeyre2020} are in fairly good agreement with the calculations \cite{McMinis2015}, which indicates a possible transition of the dielectric molecular structure with the C2/c symmetry to the metallic state with the orthorhombic structure Cmca-12 of the molecular crystalline hydrogen. The metallic character of the conductivity of the Cmca-12 structure is also shown in \cite{Zhang2018,Azadi2013,Dogan2021}. The absence of hysteresis in the pressure dependence of the direct gap is an additional indication to the molecular structure of the crystalline hydrogen. Thus, according to the conclusions of \cite{Loubeyre2020}, the experimentally observed formation of the metallic state of crystalline hydrogen is not the result of a first-order phase transition, but occurs as a result of a transition between molecular structures without a change in density.

The change in the structure of molecular crystalline hydrogen under compression along the 100\,K isotherm is studied in \cite{Norman2019,Saitov2019,Saitov2020} within the framework of the quantum molecular dynamics method. Compression of hydrogen is considered starting from a density value $\rho=1.14$\,g/cm$^3$ and corresponding pressure of 302\,GPa. The formation of a orthorhombic structure of molecular hydrogen with the Cmca-4 symmetry is observed along the molecular dynamics trajectory at $\rho=1.47$\,g/cm$^3$. This phase of the crystalline hydrogen is conductive.

The structure of the metallic molecular hydrogen is orthorhombic with the Cmca symmetry with 4 atoms per unit cell, as it is predicted in \cite{Rillo2018,Cudazzo2008,Lebegue2012,Pickard2012,Goncharov2013}. The Cmca-4 structure is stable at pressures above 440\,GPa. In \cite{Norman2019,Saitov2019,Saitov2020} it appears as a structure of metastable conducting molecular crystalline hydrogen at pressures above 544\,GPa along the 100\,K isotherm.

This work presents the results of studying the mechanism of conductivity of the crystalline molecular hydrogen structures obtained in \cite{Norman2019,Saitov2019,Saitov2020}: the monoclinic structure with C2/c symmetry and the orthorhombic structure with Cmca-4 symmetry. The Sec. II describes the theoretical method based on the quantum molecular dynamics simulation within the framework of the DFT approach. Equation of state and structural properties of solid hydrogen in the pressure range $P=302-626$\,GPa are considered in Sec. III. An analysis of the dependence of the direct and indirect gaps at the Fermi level between the conduction and valence bands for semiconducting and semimetallic states is presented in Sec. IV. Metallic states, which correspond to the metatstable Cmca-4 structure, are considered in Sec. V. Discussion of the results is presented in Sec. VI.

\section{Calculation method}
The quantum molecular dynamics method is applied within the framework of the DFT using VASP package \cite{Kresse1993,Kresse1994,Kresse1996,KresseFurthmueller1996}. Calculations are carried out in the range of hydrogen densities from 1.14\,g/cm$^3$ to 1.562\,g/cm$^3$. The number of particles in the computational cell under periodic boundary conditions is 192. The initial configuration is a monoclinic structure of the C2/c symmetry consisting 12 hydrogen atoms in the unit cell. The parameters of this cell are taken from \cite{Pickard2007}.

The calculation method contains two stages. At the first stage, the particle trajectories are calculated using the molecular dynamics method. The forces acting on ions are evaluated via the Hellmann-Feynman theorem. Then the classical Newtonian equations of motion are integrated to obtain trajectories of particles. Depending on the particles density in the computational cell, the trajectories have 10,000--20,000 steps with the timestep 0.5\,fs. During the first 1000 steps, the system is brought to the equilibrium state. On the equilibrium part of the trajectory, the proton-proton pair correlation functions (PCF) and pressure are calculated at each time step. The results are averaged over a set of equilibrium ionic configurations. Within the ionic trajectory from 1000 to 2000 statistically independent configurations are chosen for averaging.

Calculations are performed in the canonical ensemble. The ion temperature is controlled by the Nos\'{e}-Hoover thermostat \cite{Nose1984, Hoover1985}. The electron temperature coincides with the ion temperature and is established by the Fermi-Dirac distribution function, which determines the occupancies of electronic states. The temperature of the system considered is $T=100$\,K.

Wavefunctions, i.e., electron orbitals obtained by solving the Kohn–Sham equations, and the corresponding energy levels determine the ground state of the considered system for a given
configuration of atoms and temperature. The solution of the Kohn-Sham equations is found as sum of plane waves. The energy cutoff of the plane waves basis set is 1200\,eV. The electron-ion interaction is described by the potential of the projected augmented waves. The generalized gradient approximation for the exchange and correlation part of the density functional is used. The parametrisation of the functional is PBE \cite{Perdew1996}. The number of \textbf{k}-points in the Brillouin zone is chosen equal to 27. The chosen number of the k-points and the value of the plane wave basis set cutoff provide the convergence of the pressure.

The initial density of the considered system is 1.14\,g/cm$^3$. Subsequent densities are obtained by isotropic compression of the initial cell. The equilibrium structures discussed below are formed in the process of relaxation of the molecular dynamics trajectory at the constant volume and temperature. Symmetry groups and unit cells of time-averaged structures are determined using the FINDSYM code \cite{Stokes2005}.

At the second stage, the band structure is calculated for the obtained latices. Two latices of crystalline molecular hydrogen are considered: monoclinic with the C2/c consisting 12 atoms per unit cell and orthorhombic with the Cmca symmetry with 4 atoms per unit cell. The band structure calculations are carried out for the cells with 12 atoms in for the C2/c latice and 8 atoms in the latice with the Cmca symmetry. The HSE hybrid functional \cite{Heyd2003} is used, the exchange part of which contains three quarters of the PBE exchange and a quarter of the Hartry–Fock exchange, which provides more accurate evaluation of the gap between the valence and conduction bands. A Monkhorst-Pack set of $12\times12\times12$ $\bf{k}$-points in the Brilloiun zone \cite{Monkhorst1976} is specified, which allows to reach the convergence of the calculation results.

\section{Equation of state and structural properties}
\begin{figure}
\includegraphics[width=1\linewidth]{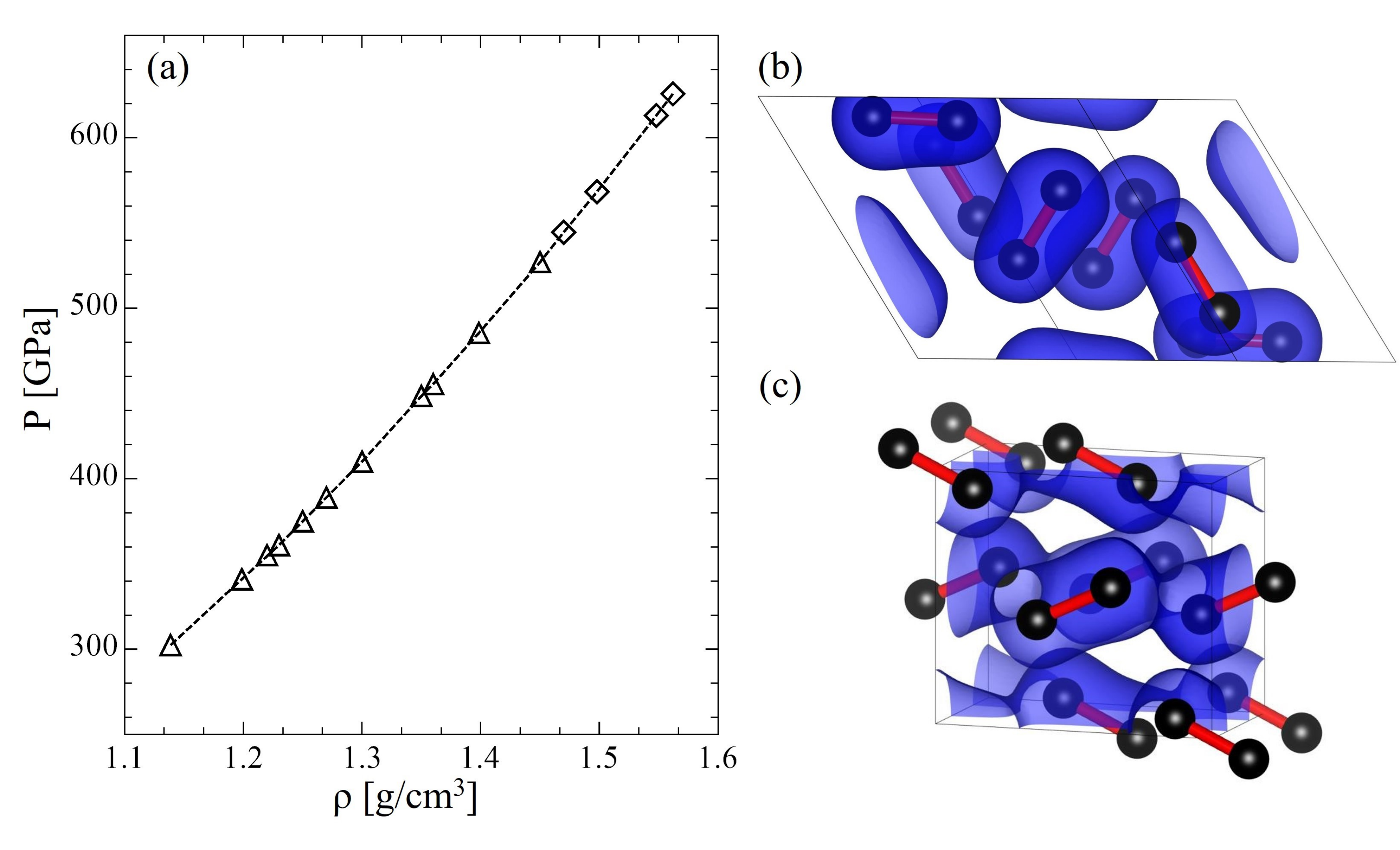}
\caption{(Color online) The 100\,K isotherm of the molecular crystalline hydrogen is shown in figure (a): triangles correspond to the monoclinic structure C2/c, diamonds correspond to the orthorhombic structure Cmca-4. Spatial arrangement of hydrogen atoms in unit cells of structures C2/c and Cmca-4 are shown in figures (b) and (c) respectively. Blue isosurfaces correspond to the electron density at the value of 0.8\,${\AA}^{-3}$.}
\label{eos}
\end{figure}

The result of calculating the dependence of pressure $P$ on density $\rho$ at a temperature of 100\,K is shown in Fig.\,\ref{eos}(a). In the density range from 1.14 to 1.45\,g/cm$^3$, crystalline molecular hydrogen retains the monoclinic structure C2/c. Points corresponding to this structure are indicated by triangles. At the density $\rho=1.47$\,g/cm$^3$, the structure changes from monoclinic to orthorhombic Cmca with 4 atoms per unit cell (diamonds in Fig.\,\ref{eos}(a)). It should be noted that this transition is only structural, since there is no density jump.

The spatial arrangement of hydrogen atoms in the unit cells of molecular phases with the C2/c and the Cmca-4 symmetries at density values 1.14\,g/cm$^3$ and 1.562\,g/cm$^3$ are shown in Figs.\,\ref{eos}(b) and \ref{eos}(c) respectively. The molecular character of the structures is indicated by the shape of the isosurface of the electron density distribution, in Figs.\,\ref{eos}(b) and \ref{eos}(c) corresponding to the value of 0.8\,${\AA}^{-3}$. Under compression and transition to the Cmca structure, crystalline hydrogen remains molecular, while the electron density in the space between molecules increases and an overlap is observed at the level of 0.8\,\AA$^{-3}$, as can be seen from Fig.\,\ref{eos}(c).
\begin{figure}
\includegraphics[width=1\linewidth]{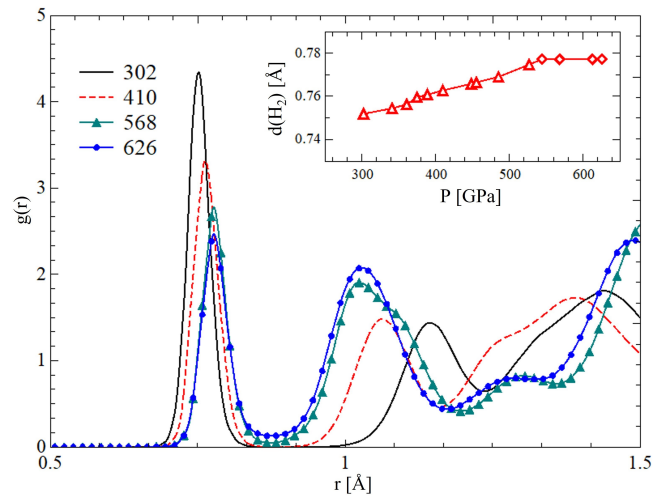}
\caption{(Color online) Pair correlation functions of molecular crystalline hydrogen at different pressures. The dependence of distance of the first maximum of the PCF $d(\rm{H_2})$ on pressure is given in the Inset.}
\label{pcf}
\end{figure}

The PCFs corresponding to the obtained structures at various pressures are shown in Fig.\,\ref{pcf}. The first peak of the PCF at the pressure of 302\,GPa is located at the distance $d(\rm{H_2})=0.75$\,\AA, which is close to the characteristic interatomic separation in the hydrogen molecule of $d(\rm{H_2})=0.74$\,\AA\,in the ground state. Compression of the initial C2/c structure results in a smooth increase in the distance corresponding to the first maximum of the PCF. As one can see from the inset in Fig.\,\ref{pcf}, the value reaches  $d(\rm{H_2})=0.78$\,\AA\,upon the formation of the Cmca structure at a pressure of 544\,GPa and then remains unchanged upon compression up to $P=626$\,GPa. Thus, upon compression of a monoclinic structure with the C2/c symmetry, an insignificant increase in the interatomic distance in the hydrogen molecule from 0.75\,\AA\,to 0.78\,\AA\,is observed. In the orthorhombic structure Cmca, this value remains unchanged with increasing pressure.

\section{Semiconducting and semimetallic states}
In order to analyze the nature of the conductivity of the considered structures of the molecular crystalline hydrogen, the band structure is calculated. To discuss the results presented below, it is necessary to clarify the concepts of direct and indirect gaps between the valence and conduction bands. The indirect bandgap $\Delta_I$ can be determined as a minimum of the difference between the boundaries of the conduction $E_c(\bf{k})$ and valence $E_v(\bf{k})$ bands, provided that the minimum of the spectrum of the conduction band and the maximum of the valence band correspond to different $\bf{k}$-vectors in the Brillouin zone. Thus, the value $\Delta_I$ can be expressed in the following form
\begin{equation}
\label{InDG}
\Delta_I = \min_{i\neq j}(E_c({\bf{k}}_i) - E_v({\bf{k}}_j)).
\end{equation}
In this case, the transition of an electron from the conduction band to the valence band is accompanied by a change in momentum $\hbar\Delta{\bf{k}}=\hbar({\bf{k}}_i - {\bf{k}}_j)$. In the process of energy absorption, in addition to the electron and photon, a phonon participates, which takes an additional momentum.

If the transition occurs without a change in momentum, which is equivalent to the case when the minimum of the spectrum of the conduction band and the maximum of the spectrum of the valence band correspond to the same $\bf{k}$-vector in the Brillouin zone, the gap is a direct one. The direct bandgap $\Delta_{D}$ can be expressed as
\begin{equation}
\label{DG}
\Delta_D = \min_{i=j}(E_c({\bf{k}}_i) - E_v({\bf{k}}_j)).
\end{equation}
For a semiconductor, both gaps are nonzero: $\Delta_{D}\neq 0$ and $\Delta_{I}\neq 0$. The condition $\Delta_{I}<\Delta_{D}$ is valid for the indirect-gap-semiconductor. Otherwise $\Delta_{I}>\Delta_{D}$, the semiconductor is direct-gap one.

In metals and semimetals, there is no indirect bandgap, i.e. the condition $\Delta_{I}=0$ is always satisfied. In metals, there is also no direct gap $\Delta_{D}=0$. Semimetals are characterized by the presence of a direct gap ($\Delta_{D}\neq 0$) between the conduction and valence bands in all directions of the $\bf{k}$-vector in the Brillouin zone. Thus, there is an overlap of the valence band in one or more sectors of the $\bf{k}$-vector with the conduction band in one or more other sectors of the $\bf{k}$-vector. As a result, the electrons that are in the valence band above the Fermi level flow into the region of the conduction band that are below the Fermi level. The resulting hole "pocket" or "pockets" in the valence band and an electronic "pocket" or "pockets" in the conduction band are signs of a semimetal. This type of the band structure of semimetals distinguishes them from metals, where the valence band is completely filled \cite{Dorofeev1984,ziman_1972,ashcroft1976solid,Volkov1983,Falkovskii1986,Falkovskii1986a,Wan2011,Weng2015,Shekhar2015}. The volumes of the electron and hole “pockets” determine the value of the electrical conductivity of the semimetal \cite{ziman_1972}.

\begin{figure}
\includegraphics[width=1\linewidth]{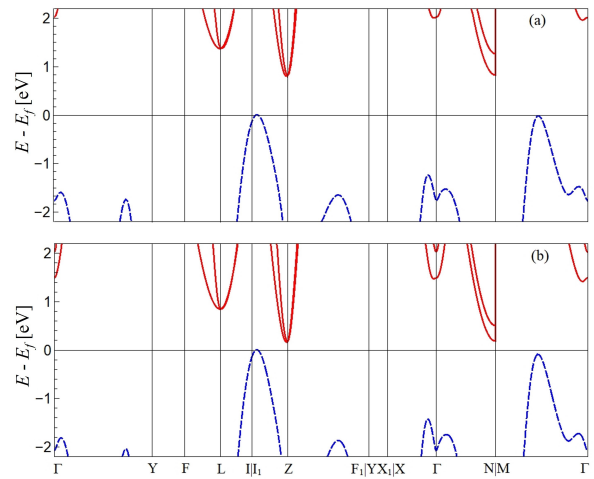}
\caption{(Color online) The band structure at pressures 302\,GPa (a) and 355\,GPa (b). The red solid lines correspond to the conduction bands, the blue dashed lines correspond to the valence bands.}
\label{bs1}
\end{figure}

As one can see from Fig.\,\ref{eos}, the monoclinic structure C2/c is stable in pressure range from 302 to 544\,GPa. The band structure for this lattice at pressures of 302 and 355\,GPa are shown in Figs.\,\ref{bs1} (a) and (b), respectively. The values of energy $E$ are measured from the Fermi energy $E_f$. The profile of the band structure has a gap between the valence and conduction bands. The values of the direct and indirect gaps at pressure of 302\,GPa are equal to $\Delta_{I}=0.802$\,eV and $\Delta_{D}=2.96$\,eV (at the k-point close to $Z$-point). The size of the gap noticeably exceeds the value of the temperature $kT=0.0086$\,eV, which determines the low electrical conductivity. As the pressure increases to 355\,GPa, the boundaries of the conduction and valence bands approach each other and the values of $\Delta_{I}$ and $\Delta_{D}$ decrease to 0.16 and 2.6\,eV, respectively.

It should be clarified that when calculating the band structure, 12 electron orbitals are considered, which, under the given conditions, corresponds to the solution of the Kohn-Sham equations with 7 occupied and 5 unoccupied electron orbitals. In order to evaluate the direct bandgap, the minimum difference between the spectrum of the last occupied and first unoccupied electron orbital is determined. In this case, the difference between the 8th and 7th electron orbitals are considered and the values of the direct gap is calculated as
\begin{equation}
\label{DG}
\Delta_D = \min_{i=j}(E_8({\bf{k}}_i) - E_7({\bf{k}}_j)).
\end{equation}

\begin{figure}
\includegraphics[width=1\linewidth]{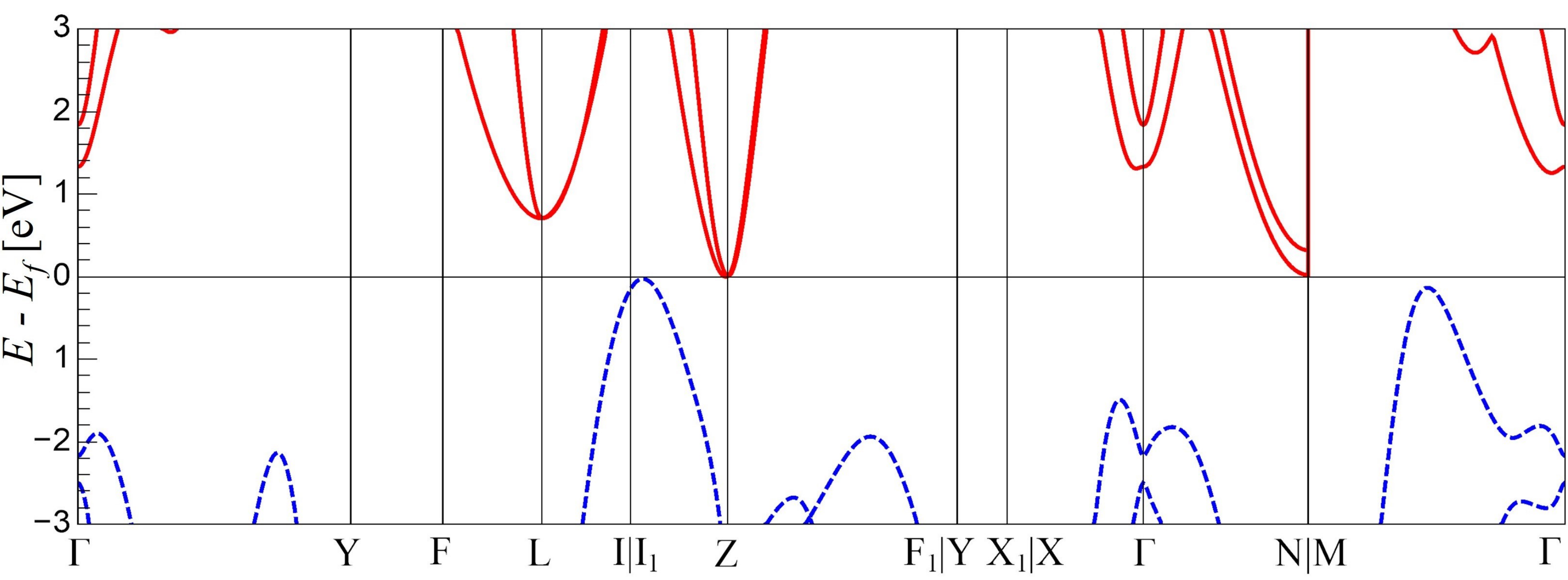}
\caption{(Color online) The band structure at pressure 361\,GPa. The symbols are similar to those shown in Fig.\,\ref{bs1}.}
\label{bs2}
\end{figure}

As it can be seen from Fig.\,\ref{bs2}, compression to a pressure of 361\,GPa leads to the closure of the indirect gap. In this case, the value of the direct gap decreases to 2.5\,eV. Thus, in the pressure range $P=302-361$\,GPa, crystalline hydrogen with the C2/c structure is an indirect-gap semiconductor.

\begin{figure}
\includegraphics[width=1\linewidth]{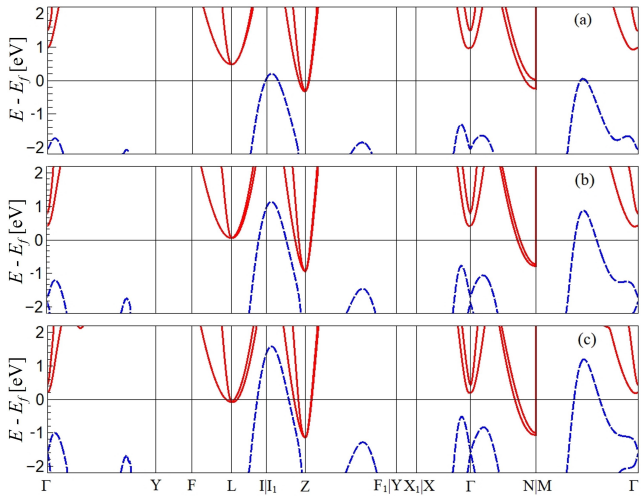}
\caption{(Color online) The band structure at pressures 389\,GPa(a), 485\,GPa (b) and 527\,GPa (c). The symbols are similar to those shown in Fig.\,\ref{bs1}.}
\label{bs3}
\end{figure}

The calculated band structures at pressures of 389, 485, and 527\,GPa are shown in Figs.\,\ref{bs3} (a), (b) and (c), respectively. Further compression to pressures above 361\,GPa leads to the overlap of the conduction and valence bands. In this case, the valence bands become partially unoccupied and, therefore, the conduction bands turns out to be partially occupied. As one can see from Fig.\,\ref{bs3}(c), while the pressure increases to 527\,GPa, the band overlap increases to 2.7\,eV, and the direct gap decreases to 0.654\,eV. As it is mentioned earlier, the resulting band structure profile is characteristic for a semimetal. Therefore, in the pressure range $P=361-527$\,GPa, the C2/c structure is a semimetal. It should be noted that the calculated value of the formation pressure of the semimetallic state of 361\,GPa is rather close to the corresponding experimental value of 350\,GPa \cite{Eremets2019}. Thus, the crossover from the semiconductor to the semimetal is observed within the C2/c structure of the molecular crystalline hydrogen.

\section{Metallic molecular crystalline hydrogen}
At the density of 1.47\,g/cm$^3$ and corresponding pressure of 544\,GPa, the structure of molecular hydrogen changes from monoclinic to orthorhombic with the Cmca symmetry group. This structure of the molecular hydrogen is metastable and exists, according to the results of \cite{Saitov2019,Saitov2020}, at the temperature of 100\,K in the pressure range from 544 to 626\,GPa. In this pressure range, the electrical conductivity is noticeably higher in comparison with the C2/c and increases up to 1200\,$(Ohm\cdot cm)^{-1}$ at the pressure of 626\,GPa.
\begin{figure}
\includegraphics[width=1\linewidth]{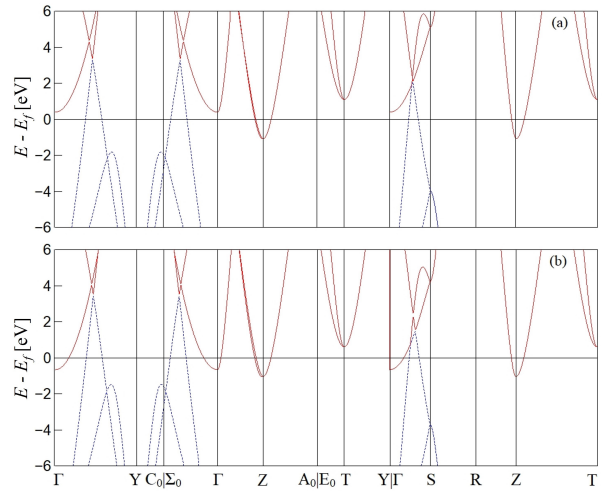}
\caption{(Color online) The band structure at pressures 544\,GPa (a) and 626\,GPa (b). The symbols are similar to those shown in Fig.\,\ref{bs1}.}
\label{bs4}
\end{figure}

As it can be seen from Fig.\,\ref{bs4}, in the Cmca-4 structure at pressures of 544\,GPa and 626\,GPa, an overlap of the valence and conduction bands occurs. This results is similar to the case of a monoclinic structure with C2/c symmetry in the pressure range from 361 to 527\,GPa. However, in comparison with the C2/c structure in the Cmca-4 the direct gap closes. Therefore, this result is an indication of the metallic conductivity of the crystalline molecular hydrogen with the Cmca-4 structure. Formation of the metallic state is associated with the structure transformation without the density discontinuity.

\section{Discussion}
\begin{figure}
\includegraphics[width=1\linewidth]{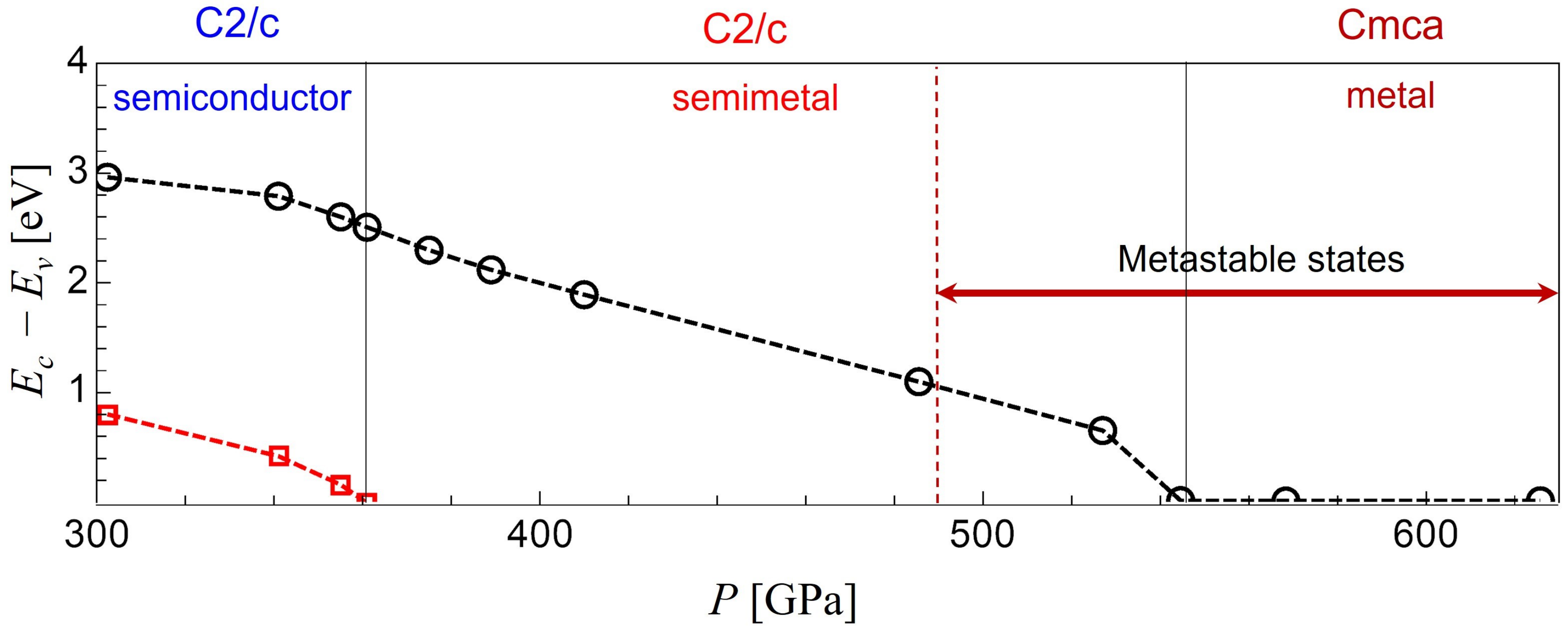}
\caption{(Color online) Pressure dependences of the direct gap (black circles connected by the dashed line) and of the indirect gap (red squares connected by the dashed line). Black vertical lines divide areas of semiconductor, semimetal, and metal. The red vertical dashed line is the border of the region of metastable states, according to the estimate \cite{Saitov2019,Saitov2020}}
\label{scheme}
\end{figure}

In Fig.\,\ref{scheme}, the results of calculating the band structure and dependencies of the magnitude of the direct and indirect gaps on pressure in the range $P=302-626$\,GPa are presented. Analysis of the results allows to distinguish three regions with different types of conductivity. A monoclinic structure with the C2/c symmetry at pressures $P=302-361$\,GPa is an indirect-gap semiconductor. At pressures above 361\,GPa, a semimetallic state of the C2/c structure is observed. The direct gap closes at $P=544$\,GPa and the formation of the Cmca-4 is observed. It indicates the metallic nature of the conductivity of the Cmca structure in the range $P=544-626$\,GPa.

The obtained pressure 361\,GPa of the formation of the semimetallic state is close to the experimentally measured value of 350\,GPa \cite{Eremets2019}.

The observed formation of the metallic Cmca-4 structure is in a good agreement with the theoretical predictions \cite{Zhang2018,Rillo2018,Azadi2014,McMinis2015,Cudazzo2008,Lebegue2012,Pickard2012,Azadi2013,Goncharov2013,Drummond2015,Monserrat2016,Azadi2017,Azadi2018}. It should be noted that, during the molecular dynamic modeling, the formation of the metallic orthorhombic structure Cmca-12, considered in \cite{Cudazzo2008,Lebegue2012,Pickard2012,Azadi2013,Goncharov2013,Drummond2015,Azadi2017,Monserrat2016}, is not observed.

According to the estimates obtained in \cite{Saitov2019,Saitov2020}, the structures arising at pressures above 487\,GPa are metastable. Therefore, the semimetallic state with the C2/c symmetry is metastable in the pressure range from 487 to 544\,GPa. The molecular metallic phase Cmca, which exists according to the calculation results in the pressure range $P=544-626$\,GPa, is completely located in the metastable region. 

Since only the stable states are observed in the experiments \cite{Dias2017,Eremets2019,Loubeyre2020}, the above mentioned result allows to suggest, that the formation of the metallic state corresponds to the first order phase transition from the stable semimetallic state with the C2/c structure to the stable atomic lattice. This suggestion excludes the observation of the metallic molecular crystalline hydrogen, assumed in the experiment \cite{Loubeyre2020}, and corresponds to the initial prediction \cite{WignerHuntington1935} and the experimental data \cite{Dias2017}.

\section{Conclusions}
The structural changes and the character of the conductivity of the molecular crystalline hydrogen are considered within the framework of the quantum molecular dynamics based on the DFT. The pressure, the pair correlation functions of protons and the band structure have been calculated at the temperature 100\,K in the pressure range $P=302-626$\,GPa. The following conclusions
are made.

1. Two structures are obtained. The first one is the monoclinic lattice with the C2/c symmetry with 12 atoms per unit cell at pressures $P=302-544$\,GPa. The second one is the orthorhombic lattice with Cmca symmetry with 4 atoms per unit cell at pressures $P=544-626$\,GPa.

2. Compression of the molecular crystal with C2/c symmetry leads to the increase of the interatomic separation in hydrogen molecule from 0.75 to 0.78\,\AA. In the orthorhombic structure Cmca-4 this parameter remains constant at the value 0.78\,\AA\,under compression up to 626\,GPa.

3. The molecular crystal with C2/c symmetry is semiconductor in the pressure range $P=302-361$\,GPa

4. The crossover from indirect-gap semiconductor state to the semimetal is observed in the C2/c lattice at the pressure 361\,GPa, which is in good agreement with experimental pressure 350\,GPa of formation of the semimetallic state \cite{Eremets2019}.

5. The crossover from the semimetallic C2/c structure to the metallic Cmca-4 structure is found at the pressure 544\,GPa. Therefore, the results obtained show that the transition from the semiconducting molecular crystalline hydrogen to the metallic state goes through the intermediate semimetallic state.

\begin{acknowledgments}
This work was supported by the Russian Science Foundation, project no. 18-19-00734. The calculations were performed in the supercomputer centers of JIHT RAS, MIPT, NRU HSE, and on the K-100 cluster Keldysh Institute of Applied Mathematics, Russian Academy of Sciences.
\end{acknowledgments}


\begin{thebibliography}{0}%
\makeatletter
\providecommand \@ifxundefined [1]{%
 \@ifx{#1\undefined}
}%
\providecommand \@ifnum [1]{%
 \ifnum #1\expandafter \@firstoftwo
 \else \expandafter \@secondoftwo
 \fi
}%
\providecommand \@ifx [1]{%
 \ifx #1\expandafter \@firstoftwo
 \else \expandafter \@secondoftwo
 \fi
}%
\providecommand \natexlab [1]{#1}%
\providecommand \enquote  [1]{``#1''}%
\providecommand \bibnamefont  [1]{#1}%
\providecommand \bibfnamefont [1]{#1}%
\providecommand \citenamefont [1]{#1}%
\providecommand \href@noop [0]{\@secondoftwo}%
\providecommand \href [0]{\begingroup \@sanitize@url \@href}%
\providecommand \@href[1]{\@@startlink{#1}\@@href}%
\providecommand \@@href[1]{\endgroup#1\@@endlink}%
\providecommand \@sanitize@url [0]{\catcode `\\12\catcode `\$12\catcode
  `\&12\catcode `\#12\catcode `\^12\catcode `\_12\catcode `\%12\relax}%
\providecommand \@@startlink[1]{}%
\providecommand \@@endlink[0]{}%
\providecommand \url  [0]{\begingroup\@sanitize@url \@url }%
\providecommand \@url [1]{\endgroup\@href {#1}{\urlprefix }}%
\providecommand \urlprefix  [0]{URL }%
\providecommand \Eprint [0]{\href }%
\providecommand \doibase [0]{https://doi.org/}%
\providecommand \selectlanguage [0]{\@gobble}%
\providecommand \bibinfo  [0]{\@secondoftwo}%
\providecommand \bibfield  [0]{\@secondoftwo}%
\providecommand \translation [1]{[#1]}%
\providecommand \BibitemOpen [0]{}%
\providecommand \bibitemStop [0]{}%
\providecommand \bibitemNoStop [0]{.\EOS\space}%
\providecommand \EOS [0]{\spacefactor3000\relax}%
\providecommand \BibitemShut  [1]{\csname bibitem#1\endcsname}%
\let\auto@bib@innerbib\@empty
\end{thebibliography}%


\begin{thebibliography}{58}%
\makeatletter
\providecommand \@ifxundefined [1]{%
 \@ifx{#1\undefined}
}%
\providecommand \@ifnum [1]{%
 \ifnum #1\expandafter \@firstoftwo
 \else \expandafter \@secondoftwo
 \fi
}%
\providecommand \@ifx [1]{%
 \ifx #1\expandafter \@firstoftwo
 \else \expandafter \@secondoftwo
 \fi
}%
\providecommand \natexlab [1]{#1}%
\providecommand \enquote  [1]{``#1''}%
\providecommand \bibnamefont  [1]{#1}%
\providecommand \bibfnamefont [1]{#1}%
\providecommand \citenamefont [1]{#1}%
\providecommand \href@noop [0]{\@secondoftwo}%
\providecommand \href [0]{\begingroup \@sanitize@url \@href}%
\providecommand \@href[1]{\@@startlink{#1}\@@href}%
\providecommand \@@href[1]{\endgroup#1\@@endlink}%
\providecommand \@sanitize@url [0]{\catcode `\\12\catcode `\$12\catcode
  `\&12\catcode `\#12\catcode `\^12\catcode `\_12\catcode `\%12\relax}%
\providecommand \@@startlink[1]{}%
\providecommand \@@endlink[0]{}%
\providecommand \url  [0]{\begingroup\@sanitize@url \@url }%
\providecommand \@url [1]{\endgroup\@href {#1}{\urlprefix }}%
\providecommand \urlprefix  [0]{URL }%
\providecommand \Eprint [0]{\href }%
\providecommand \doibase [0]{https://doi.org/}%
\providecommand \selectlanguage [0]{\@gobble}%
\providecommand \bibinfo  [0]{\@secondoftwo}%
\providecommand \bibfield  [0]{\@secondoftwo}%
\providecommand \translation [1]{[#1]}%
\providecommand \BibitemOpen [0]{}%
\providecommand \bibitemStop [0]{}%
\providecommand \bibitemNoStop [0]{.\EOS\space}%
\providecommand \EOS [0]{\spacefactor3000\relax}%
\providecommand \BibitemShut  [1]{\csname bibitem#1\endcsname}%
\let\auto@bib@innerbib\@empty
\bibitem [{\citenamefont {Wigner}\ and\ \citenamefont
  {Huntington}(1935)}]{WignerHuntington1935}%
  \BibitemOpen
  \bibfield  {author} {\bibinfo {author} {\bibfnamefont {E.}~\bibnamefont
  {Wigner}}\ and\ \bibinfo {author} {\bibfnamefont {H.~B.}\ \bibnamefont
  {Huntington}},\ }\href {https://doi.org/10.1063/1.1749590} {\bibfield
  {journal} {\bibinfo  {journal} {J. Chem. Phys.}\ }\textbf {\bibinfo {volume}
  {3}},\ \bibinfo {pages} {764} (\bibinfo {year} {1935})}\BibitemShut {NoStop}%
\bibitem [{\citenamefont {Ashcroft}(1968)}]{Ashcroft1968}%
  \BibitemOpen
  \bibfield  {author} {\bibinfo {author} {\bibfnamefont {N.~W.}\ \bibnamefont
  {Ashcroft}},\ }\href {https://doi.org/10.1103/PhysRevLett.21.1748} {\bibfield
   {journal} {\bibinfo  {journal} {Phys. Rev. Lett.}\ }\textbf {\bibinfo
  {volume} {21}},\ \bibinfo {pages} {1748} (\bibinfo {year}
  {1968})}\BibitemShut {NoStop}%
\bibitem [{\citenamefont {Kagan}\ and\ \citenamefont
  {Brovman}(1972)}]{Kagan1972}%
  \BibitemOpen
  \bibfield  {author} {\bibinfo {author} {\bibfnamefont {Y.~M.}\ \bibnamefont
  {Kagan}}\ and\ \bibinfo {author} {\bibfnamefont {E.~G.}\ \bibnamefont
  {Brovman}},\ }\href {https://doi.org/10.1070/PU1972v014n06ABEH004828}
  {\bibfield  {journal} {\bibinfo  {journal} {Soviet Physics Uspekhi}\ }\textbf
  {\bibinfo {volume} {14}},\ \bibinfo {pages} {809} (\bibinfo {year}
  {1972})}\BibitemShut {NoStop}%
\bibitem [{\citenamefont {Brovman}\ \emph
  {et~al.}(1972{\natexlab{a}})\citenamefont {Brovman}, \citenamefont {Kagan},\
  and\ \citenamefont {Kholas}}]{Brovman1972}%
  \BibitemOpen
  \bibfield  {author} {\bibinfo {author} {\bibfnamefont {E.~G.}\ \bibnamefont
  {Brovman}}, \bibinfo {author} {\bibfnamefont {Y.}~\bibnamefont {Kagan}},\
  and\ \bibinfo {author} {\bibfnamefont {A.}~\bibnamefont {Kholas}},\ }\href
  {https://ui.adsabs.harvard.edu/abs/1972JETP...35..783B} {\bibfield  {journal}
  {\bibinfo  {journal} {Sov. J. Exp. Theor. Phys.}\ }\textbf {\bibinfo {volume}
  {35}},\ \bibinfo {pages} {783} (\bibinfo {year}
  {1972}{\natexlab{a}})}\BibitemShut {NoStop}%
\bibitem [{\citenamefont {Brovman}\ \emph
  {et~al.}(1972{\natexlab{b}})\citenamefont {Brovman}, \citenamefont {Kagan},\
  and\ \citenamefont {Kholas}}]{Brovman1972a}%
  \BibitemOpen
  \bibfield  {author} {\bibinfo {author} {\bibfnamefont {E.~G.}\ \bibnamefont
  {Brovman}}, \bibinfo {author} {\bibfnamefont {Y.}~\bibnamefont {Kagan}},\
  and\ \bibinfo {author} {\bibfnamefont {A.}~\bibnamefont {Kholas}},\ }\href
  {https://ui.adsabs.harvard.edu/abs/1972JETP...34.1300B} {\bibfield  {journal}
  {\bibinfo  {journal} {Sov. J. Exp. Theor. Phys.}\ }\textbf {\bibinfo {volume}
  {34}},\ \bibinfo {pages} {1300} (\bibinfo {year}
  {1972}{\natexlab{b}})}\BibitemShut {NoStop}%
\bibitem [{\citenamefont {Brovman}\ \emph {et~al.}(1973)\citenamefont
  {Brovman}, \citenamefont {Kagan}, \citenamefont {Kholas},\ and\ \citenamefont
  {Pushkarev}}]{Brovman1973}%
  \BibitemOpen
  \bibfield  {author} {\bibinfo {author} {\bibfnamefont {E.~G.}\ \bibnamefont
  {Brovman}}, \bibinfo {author} {\bibfnamefont {Y.}~\bibnamefont {Kagan}},
  \bibinfo {author} {\bibfnamefont {A.}~\bibnamefont {Kholas}},\ and\ \bibinfo
  {author} {\bibfnamefont {V.~V.}\ \bibnamefont {Pushkarev}},\ }\href
  {https://ui.adsabs.harvard.edu/abs/1973JETPL..18..160B} {\bibfield  {journal}
  {\bibinfo  {journal} {Sov. J. Exp. Theor. Phys.}\ }\textbf {\bibinfo {volume}
  {18}},\ \bibinfo {pages} {160} (\bibinfo {year} {1973})}\BibitemShut
  {NoStop}%
\bibitem [{\citenamefont {Kagan}\ \emph {et~al.}(1977)\citenamefont {Kagan},
  \citenamefont {Pushkarev},\ and\ \citenamefont {Kholas}}]{Kagan1977}%
  \BibitemOpen
  \bibfield  {author} {\bibinfo {author} {\bibfnamefont {Y.}~\bibnamefont
  {Kagan}}, \bibinfo {author} {\bibfnamefont {V.~V.}\ \bibnamefont
  {Pushkarev}},\ and\ \bibinfo {author} {\bibfnamefont {A.}~\bibnamefont
  {Kholas}},\ }\href {https://ui.adsabs.harvard.edu/abs/1977JETP...46..511K}
  {\bibfield  {journal} {\bibinfo  {journal} {Sov. J. Exp. Theor. Phys.}\
  }\textbf {\bibinfo {volume} {46}},\ \bibinfo {pages} {511} (\bibinfo {year}
  {1977})}\BibitemShut {NoStop}%
\bibitem [{\citenamefont {Pickard}\ and\ \citenamefont
  {Needs}(2007)}]{Pickard2007}%
  \BibitemOpen
  \bibfield  {author} {\bibinfo {author} {\bibfnamefont {C.~J.}\ \bibnamefont
  {Pickard}}\ and\ \bibinfo {author} {\bibfnamefont {R.~J.}\ \bibnamefont
  {Needs}},\ }\href {https://doi.org/10.1038/nphys625} {\bibfield  {journal}
  {\bibinfo  {journal} {Nat. Phys.}\ }\textbf {\bibinfo {volume} {3}},\
  \bibinfo {pages} {473} (\bibinfo {year} {2007})}\BibitemShut {NoStop}%
\bibitem [{\citenamefont {McMahon}\ and\ \citenamefont
  {Ceperley}(2011)}]{McMahon2011}%
  \BibitemOpen
  \bibfield  {author} {\bibinfo {author} {\bibfnamefont {J.~M.}\ \bibnamefont
  {McMahon}}\ and\ \bibinfo {author} {\bibfnamefont {D.~M.}\ \bibnamefont
  {Ceperley}},\ }\href {https://doi.org/10.1103/PhysRevLett.106.165302}
  {\bibfield  {journal} {\bibinfo  {journal} {Phys. Rev. Lett.}\ }\textbf
  {\bibinfo {volume} {106}},\ \bibinfo {eid} {165302} (\bibinfo {year}
  {2011})},\ \Eprint {https://arxiv.org/abs/1011.5028} {arXiv:1011.5028
  [cond-mat.mtrl-sci]} \BibitemShut {NoStop}%
\bibitem [{\citenamefont {Degtyarenko}\ and\ \citenamefont
  {Mazur}(2016)}]{Degtyarenko2016}%
  \BibitemOpen
  \bibfield  {author} {\bibinfo {author} {\bibfnamefont {N.~N.}\ \bibnamefont
  {Degtyarenko}}\ and\ \bibinfo {author} {\bibfnamefont {E.~A.}\ \bibnamefont
  {Mazur}},\ }\href {https://doi.org/10.1134/S0021364016170069} {\bibfield
  {journal} {\bibinfo  {journal} {J. Exp. Theor. Phys. Lett.}\ }\textbf
  {\bibinfo {volume} {104}},\ \bibinfo {pages} {319} (\bibinfo {year}
  {2016})}\BibitemShut {NoStop}%
\bibitem [{\citenamefont {Kudryashov}\ \emph {et~al.}(2016)\citenamefont
  {Kudryashov}, \citenamefont {Kutukov},\ and\ \citenamefont
  {Mazur}}]{Kudryashov2016}%
  \BibitemOpen
  \bibfield  {author} {\bibinfo {author} {\bibfnamefont {N.~A.}\ \bibnamefont
  {Kudryashov}}, \bibinfo {author} {\bibfnamefont {A.~A.}\ \bibnamefont
  {Kutukov}},\ and\ \bibinfo {author} {\bibfnamefont {E.~A.}\ \bibnamefont
  {Mazur}},\ }\href {https://doi.org/10.1134/S0021364016190061} {\bibfield
  {journal} {\bibinfo  {journal} {J. Exp. Theor. Phys. Lett.}\ }\textbf
  {\bibinfo {volume} {104}},\ \bibinfo {pages} {460} (\bibinfo {year}
  {2016})}\BibitemShut {NoStop}%
\bibitem [{\citenamefont {Degtyarenko}\ \emph {et~al.}(2017)\citenamefont
  {Degtyarenko}, \citenamefont {Mazur},\ and\ \citenamefont
  {Grishakov}}]{Degtyarenko2017}%
  \BibitemOpen
  \bibfield  {author} {\bibinfo {author} {\bibfnamefont {N.~N.}\ \bibnamefont
  {Degtyarenko}}, \bibinfo {author} {\bibfnamefont {E.~A.}\ \bibnamefont
  {Mazur}},\ and\ \bibinfo {author} {\bibfnamefont {K.~S.}\ \bibnamefont
  {Grishakov}},\ }\href {https://doi.org/10.1134/S0021364017100058} {\bibfield
  {journal} {\bibinfo  {journal} {J. Exp. Theor. Phys. Lett.}\ }\textbf
  {\bibinfo {volume} {105}},\ \bibinfo {pages} {664} (\bibinfo {year}
  {2017})}\BibitemShut {NoStop}%
\bibitem [{\citenamefont {Kudryashov}\ \emph {et~al.}(2017)\citenamefont
  {Kudryashov}, \citenamefont {Kutukov},\ and\ \citenamefont
  {Mazur}}]{Kudryashov2017}%
  \BibitemOpen
  \bibfield  {author} {\bibinfo {author} {\bibfnamefont {N.~A.}\ \bibnamefont
  {Kudryashov}}, \bibinfo {author} {\bibfnamefont {A.~A.}\ \bibnamefont
  {Kutukov}},\ and\ \bibinfo {author} {\bibfnamefont {E.~A.}\ \bibnamefont
  {Mazur}},\ }\href {https://doi.org/10.1134/S0021364017070098} {\bibfield
  {journal} {\bibinfo  {journal} {J. Exp. Theor. Phys. Lett.}\ }\textbf
  {\bibinfo {volume} {105}},\ \bibinfo {pages} {430} (\bibinfo {year}
  {2017})}\BibitemShut {NoStop}%
\bibitem [{\citenamefont {Zhang}\ \emph {et~al.}(2018)\citenamefont {Zhang},
  \citenamefont {Wang},\ and\ \citenamefont {Li}}]{Zhang2018}%
  \BibitemOpen
  \bibfield  {author} {\bibinfo {author} {\bibfnamefont {X.-W.}\ \bibnamefont
  {Zhang}}, \bibinfo {author} {\bibfnamefont {E.-G.}\ \bibnamefont {Wang}},\
  and\ \bibinfo {author} {\bibfnamefont {X.-Z.}\ \bibnamefont {Li}},\ }\href
  {https://doi.org/10.1103/PhysRevB.98.134110} {\bibfield  {journal} {\bibinfo
  {journal} {Phys. Rev. B}\ }\textbf {\bibinfo {volume} {98}},\ \bibinfo {eid}
  {134110} (\bibinfo {year} {2018})},\ \Eprint
  {https://arxiv.org/abs/1805.01625} {arXiv:1805.01625 [cond-mat.mtrl-sci]}
  \BibitemShut {NoStop}%
\bibitem [{\citenamefont {Rillo}\ \emph {et~al.}(2018)\citenamefont {Rillo},
  \citenamefont {Morales}, \citenamefont {Ceperley},\ and\ \citenamefont
  {Pierleoni}}]{Rillo2018}%
  \BibitemOpen
  \bibfield  {author} {\bibinfo {author} {\bibfnamefont {G.}~\bibnamefont
  {Rillo}}, \bibinfo {author} {\bibfnamefont {M.~A.}\ \bibnamefont {Morales}},
  \bibinfo {author} {\bibfnamefont {D.~M.}\ \bibnamefont {Ceperley}},\ and\
  \bibinfo {author} {\bibfnamefont {C.}~\bibnamefont {Pierleoni}},\ }\href
  {https://doi.org/10.1063/1.5001387} {\bibfield  {journal} {\bibinfo
  {journal} {J. Chem. Phys.}\ }\textbf {\bibinfo {volume} {148}},\ \bibinfo
  {eid} {102314} (\bibinfo {year} {2018})},\ \Eprint
  {https://arxiv.org/abs/1708.07344} {arXiv:1708.07344 [cond-mat.mtrl-sci]}
  \BibitemShut {NoStop}%
\bibitem [{\citenamefont {Azadi}\ \emph {et~al.}(2014)\citenamefont {Azadi},
  \citenamefont {Monserrat}, \citenamefont {Foulkes},\ and\ \citenamefont
  {Needs}}]{Azadi2014}%
  \BibitemOpen
  \bibfield  {author} {\bibinfo {author} {\bibfnamefont {S.}~\bibnamefont
  {Azadi}}, \bibinfo {author} {\bibfnamefont {B.}~\bibnamefont {Monserrat}},
  \bibinfo {author} {\bibfnamefont {W.~M.~C.}\ \bibnamefont {Foulkes}},\ and\
  \bibinfo {author} {\bibfnamefont {R.~J.}\ \bibnamefont {Needs}},\ }\href
  {https://doi.org/10.1103/PhysRevLett.112.165501} {\bibfield  {journal}
  {\bibinfo  {journal} {Phys. Rev. Lett.}\ }\textbf {\bibinfo {volume} {112}},\
  \bibinfo {eid} {165501} (\bibinfo {year} {2014})},\ \Eprint
  {https://arxiv.org/abs/1403.3681} {arXiv:1403.3681 [cond-mat.mtrl-sci]}
  \BibitemShut {NoStop}%
\bibitem [{\citenamefont {McMinis}\ \emph {et~al.}(2015)\citenamefont
  {McMinis}, \citenamefont {Clay}, \citenamefont {Lee},\ and\ \citenamefont
  {Morales}}]{McMinis2015}%
  \BibitemOpen
  \bibfield  {author} {\bibinfo {author} {\bibfnamefont {J.}~\bibnamefont
  {McMinis}}, \bibinfo {author} {\bibfnamefont {R.~C.}\ \bibnamefont {Clay}},
  \bibinfo {author} {\bibfnamefont {D.}~\bibnamefont {Lee}},\ and\ \bibinfo
  {author} {\bibfnamefont {M.~A.}\ \bibnamefont {Morales}},\ }\href
  {https://doi.org/10.1103/PhysRevLett.114.105305} {\bibfield  {journal}
  {\bibinfo  {journal} {Phys. Rev. Lett.}\ }\textbf {\bibinfo {volume} {114}},\
  \bibinfo {eid} {105305} (\bibinfo {year} {2015})}\BibitemShut {NoStop}%
\bibitem [{\citenamefont {Dias}\ and\ \citenamefont
  {Silvera}(2017)}]{Dias2017}%
  \BibitemOpen
  \bibfield  {author} {\bibinfo {author} {\bibfnamefont {R.~P.}\ \bibnamefont
  {Dias}}\ and\ \bibinfo {author} {\bibfnamefont {I.~F.}\ \bibnamefont
  {Silvera}},\ }\href {https://doi.org/10.1126/science.aal1579} {\bibfield
  {journal} {\bibinfo  {journal} {Science}\ }\textbf {\bibinfo {volume}
  {355}},\ \bibinfo {pages} {715} (\bibinfo {year} {2017})},\ \Eprint
  {https://arxiv.org/abs/1610.01634} {arXiv:1610.01634 [cond-mat.mtrl-sci]}
  \BibitemShut {NoStop}%
\bibitem [{\citenamefont {Eremets}\ \emph {et~al.}(2019)\citenamefont
  {Eremets}, \citenamefont {Drozdov}, \citenamefont {Kong},\ and\ \citenamefont
  {Wang}}]{Eremets2019}%
  \BibitemOpen
  \bibfield  {author} {\bibinfo {author} {\bibfnamefont {M.~I.}\ \bibnamefont
  {Eremets}}, \bibinfo {author} {\bibfnamefont {A.~P.}\ \bibnamefont
  {Drozdov}}, \bibinfo {author} {\bibfnamefont {P.~P.}\ \bibnamefont {Kong}},\
  and\ \bibinfo {author} {\bibfnamefont {H.}~\bibnamefont {Wang}},\ }\href
  {https://doi.org/10.1038/s41567-019-0646-x} {\bibfield  {journal} {\bibinfo
  {journal} {Nat. Phys.}\ }\textbf {\bibinfo {volume} {15}},\ \bibinfo {pages}
  {1246} (\bibinfo {year} {2019})}\BibitemShut {NoStop}%
\bibitem [{\citenamefont {Loubeyre}\ \emph {et~al.}(2020)\citenamefont
  {Loubeyre}, \citenamefont {Occelli},\ and\ \citenamefont
  {Dumas}}]{Loubeyre2020}%
  \BibitemOpen
  \bibfield  {author} {\bibinfo {author} {\bibfnamefont {P.}~\bibnamefont
  {Loubeyre}}, \bibinfo {author} {\bibfnamefont {F.}~\bibnamefont {Occelli}},\
  and\ \bibinfo {author} {\bibfnamefont {P.}~\bibnamefont {Dumas}},\ }\href
  {https://doi.org/10.1038/s41586-019-1927-3} {\bibfield  {journal} {\bibinfo
  {journal} {Nature}\ }\textbf {\bibinfo {volume} {577}},\ \bibinfo {pages}
  {631} (\bibinfo {year} {2020})}\BibitemShut {NoStop}%
\bibitem [{\citenamefont {Johnson}\ and\ \citenamefont
  {Ashcroft}(2000)}]{Johnson2000}%
  \BibitemOpen
  \bibfield  {author} {\bibinfo {author} {\bibfnamefont {K.~A.}\ \bibnamefont
  {Johnson}}\ and\ \bibinfo {author} {\bibfnamefont {N.~W.}\ \bibnamefont
  {Ashcroft}},\ }\href {https://doi.org/10.1038/35001024} {\bibfield  {journal}
  {\bibinfo  {journal} {Nature}\ }\textbf {\bibinfo {volume} {403}},\ \bibinfo
  {pages} {632} (\bibinfo {year} {2000})}\BibitemShut {NoStop}%
\bibitem [{\citenamefont {Cudazzo}\ \emph {et~al.}(2008)\citenamefont
  {Cudazzo}, \citenamefont {Profeta}, \citenamefont {Sanna}, \citenamefont
  {Floris}, \citenamefont {Continenza}, \citenamefont {Massidda},\ and\
  \citenamefont {Gross}}]{Cudazzo2008}%
  \BibitemOpen
  \bibfield  {author} {\bibinfo {author} {\bibfnamefont {P.}~\bibnamefont
  {Cudazzo}}, \bibinfo {author} {\bibfnamefont {G.}~\bibnamefont {Profeta}},
  \bibinfo {author} {\bibfnamefont {A.}~\bibnamefont {Sanna}}, \bibinfo
  {author} {\bibfnamefont {A.}~\bibnamefont {Floris}}, \bibinfo {author}
  {\bibfnamefont {A.}~\bibnamefont {Continenza}}, \bibinfo {author}
  {\bibfnamefont {S.}~\bibnamefont {Massidda}},\ and\ \bibinfo {author}
  {\bibfnamefont {E.~K.~U.}\ \bibnamefont {Gross}},\ }\href
  {https://doi.org/10.1103/PhysRevLett.100.257001} {\bibfield  {journal}
  {\bibinfo  {journal} {Phys. Rev. Lett.}\ }\textbf {\bibinfo {volume} {100}},\
  \bibinfo {eid} {257001} (\bibinfo {year} {2008})}\BibitemShut {NoStop}%
\bibitem [{\citenamefont {Leb{\`e}gue}\ \emph {et~al.}(2012)\citenamefont
  {Leb{\`e}gue}, \citenamefont {Moyses~Araujo}, \citenamefont {Kim},
  \citenamefont {Ramzan}, \citenamefont {Mao},\ and\ \citenamefont
  {Ahuja}}]{Lebegue2012}%
  \BibitemOpen
  \bibfield  {author} {\bibinfo {author} {\bibfnamefont {S.}~\bibnamefont
  {Leb{\`e}gue}}, \bibinfo {author} {\bibfnamefont {C.}~\bibnamefont
  {Moyses~Araujo}}, \bibinfo {author} {\bibfnamefont {D.~Y.}\ \bibnamefont
  {Kim}}, \bibinfo {author} {\bibfnamefont {M.}~\bibnamefont {Ramzan}},
  \bibinfo {author} {\bibfnamefont {H.-k.}\ \bibnamefont {Mao}},\ and\ \bibinfo
  {author} {\bibfnamefont {R.}~\bibnamefont {Ahuja}},\ }\href
  {https://doi.org/10.1073/pnas.1207065109} {\bibfield  {journal} {\bibinfo
  {journal} {Proc. Natl. Acad. Sci. U.S.A.}\ }\textbf {\bibinfo {volume}
  {109}},\ \bibinfo {pages} {9766} (\bibinfo {year} {2012})}\BibitemShut
  {NoStop}%
\bibitem [{\citenamefont {Pickard}\ \emph {et~al.}(2012)\citenamefont
  {Pickard}, \citenamefont {Martinez-Canales},\ and\ \citenamefont
  {Needs}}]{Pickard2012}%
  \BibitemOpen
  \bibfield  {author} {\bibinfo {author} {\bibfnamefont {C.~J.}\ \bibnamefont
  {Pickard}}, \bibinfo {author} {\bibfnamefont {M.}~\bibnamefont
  {Martinez-Canales}},\ and\ \bibinfo {author} {\bibfnamefont {R.~J.}\
  \bibnamefont {Needs}},\ }\href {https://doi.org/10.1103/PhysRevB.85.214114}
  {\bibfield  {journal} {\bibinfo  {journal} {Phys. Rev. B}\ }\textbf {\bibinfo
  {volume} {85}},\ \bibinfo {eid} {214114} (\bibinfo {year} {2012})},\ \Eprint
  {https://arxiv.org/abs/1204.3304} {arXiv:1204.3304 [cond-mat.mtrl-sci]}
  \BibitemShut {NoStop}%
\bibitem [{\citenamefont {Azadi}\ \emph {et~al.}(2013)\citenamefont {Azadi},
  \citenamefont {Foulkes},\ and\ \citenamefont {K{\"u}hne}}]{Azadi2013}%
  \BibitemOpen
  \bibfield  {author} {\bibinfo {author} {\bibfnamefont {S.}~\bibnamefont
  {Azadi}}, \bibinfo {author} {\bibfnamefont {W.~M.~C.}\ \bibnamefont
  {Foulkes}},\ and\ \bibinfo {author} {\bibfnamefont {T.~D.}\ \bibnamefont
  {K{\"u}hne}},\ }\href {https://doi.org/10.1088/1367-2630/15/11/113005}
  {\bibfield  {journal} {\bibinfo  {journal} {New J. Phys.}\ }\textbf {\bibinfo
  {volume} {15}},\ \bibinfo {eid} {113005} (\bibinfo {year} {2013})},\ \Eprint
  {https://arxiv.org/abs/1307.1463} {arXiv:1307.1463 [cond-mat.mtrl-sci]}
  \BibitemShut {NoStop}%
\bibitem [{\citenamefont {Goncharov}\ \emph {et~al.}(2013)\citenamefont
  {Goncharov}, \citenamefont {Tse}, \citenamefont {Wang}, \citenamefont {Yang},
  \citenamefont {Struzhkin}, \citenamefont {Howie},\ and\ \citenamefont
  {Gregoryanz}}]{Goncharov2013}%
  \BibitemOpen
  \bibfield  {author} {\bibinfo {author} {\bibfnamefont {A.~F.}\ \bibnamefont
  {Goncharov}}, \bibinfo {author} {\bibfnamefont {J.~S.}\ \bibnamefont {Tse}},
  \bibinfo {author} {\bibfnamefont {H.}~\bibnamefont {Wang}}, \bibinfo {author}
  {\bibfnamefont {J.}~\bibnamefont {Yang}}, \bibinfo {author} {\bibfnamefont
  {V.~V.}\ \bibnamefont {Struzhkin}}, \bibinfo {author} {\bibfnamefont {R.~T.}\
  \bibnamefont {Howie}},\ and\ \bibinfo {author} {\bibfnamefont
  {E.}~\bibnamefont {Gregoryanz}},\ }\href
  {https://doi.org/10.1103/PhysRevB.87.024101} {\bibfield  {journal} {\bibinfo
  {journal} {Phys. Rev. B}\ }\textbf {\bibinfo {volume} {87}},\ \bibinfo {eid}
  {024101} (\bibinfo {year} {2013})},\ \Eprint
  {https://arxiv.org/abs/1209.3895} {arXiv:1209.3895 [cond-mat.mtrl-sci]}
  \BibitemShut {NoStop}%
\bibitem [{\citenamefont {Drummond}\ \emph {et~al.}(2015)\citenamefont
  {Drummond}, \citenamefont {Monserrat}, \citenamefont {Lloyd-Williams},
  \citenamefont {R{\'\i}os}, \citenamefont {Pickard},\ and\ \citenamefont
  {Needs}}]{Drummond2015}%
  \BibitemOpen
  \bibfield  {author} {\bibinfo {author} {\bibfnamefont {N.~D.}\ \bibnamefont
  {Drummond}}, \bibinfo {author} {\bibfnamefont {B.}~\bibnamefont {Monserrat}},
  \bibinfo {author} {\bibfnamefont {J.~H.}\ \bibnamefont {Lloyd-Williams}},
  \bibinfo {author} {\bibfnamefont {P.~L.}\ \bibnamefont {R{\'\i}os}}, \bibinfo
  {author} {\bibfnamefont {C.~J.}\ \bibnamefont {Pickard}},\ and\ \bibinfo
  {author} {\bibfnamefont {R.~J.}\ \bibnamefont {Needs}},\ }\href
  {https://doi.org/10.1038/ncomms8794} {\bibfield  {journal} {\bibinfo
  {journal} {Nat. Comm.}\ }\textbf {\bibinfo {volume} {6}},\ \bibinfo {eid}
  {7794} (\bibinfo {year} {2015})},\ \Eprint {https://arxiv.org/abs/1508.02313}
  {arXiv:1508.02313 [cond-mat.mtrl-sci]} \BibitemShut {NoStop}%
\bibitem [{\citenamefont {Monserrat}\ \emph {et~al.}(2016)\citenamefont
  {Monserrat}, \citenamefont {Needs}, \citenamefont {Gregoryanz},\ and\
  \citenamefont {Pickard}}]{Monserrat2016}%
  \BibitemOpen
  \bibfield  {author} {\bibinfo {author} {\bibfnamefont {B.}~\bibnamefont
  {Monserrat}}, \bibinfo {author} {\bibfnamefont {R.~J.}\ \bibnamefont
  {Needs}}, \bibinfo {author} {\bibfnamefont {E.}~\bibnamefont {Gregoryanz}},\
  and\ \bibinfo {author} {\bibfnamefont {C.~J.}\ \bibnamefont {Pickard}},\
  }\href {https://doi.org/10.1103/PhysRevB.94.134101} {\bibfield  {journal}
  {\bibinfo  {journal} {Phys. Rev. B}\ }\textbf {\bibinfo {volume} {94}},\
  \bibinfo {eid} {134101} (\bibinfo {year} {2016})},\ \Eprint
  {https://arxiv.org/abs/1609.07486} {arXiv:1609.07486 [cond-mat.mtrl-sci]}
  \BibitemShut {NoStop}%
\bibitem [{\citenamefont {Azadi}\ \emph {et~al.}(2017)\citenamefont {Azadi},
  \citenamefont {Drummond},\ and\ \citenamefont {Foulkes}}]{Azadi2017}%
  \BibitemOpen
  \bibfield  {author} {\bibinfo {author} {\bibfnamefont {S.}~\bibnamefont
  {Azadi}}, \bibinfo {author} {\bibfnamefont {N.~D.}\ \bibnamefont
  {Drummond}},\ and\ \bibinfo {author} {\bibfnamefont {W.~M.~C.}\ \bibnamefont
  {Foulkes}},\ }\href {https://doi.org/10.1103/PhysRevB.95.035142} {\bibfield
  {journal} {\bibinfo  {journal} {Phys. Rev. B}\ }\textbf {\bibinfo {volume}
  {95}},\ \bibinfo {eid} {035142} (\bibinfo {year} {2017})},\ \Eprint
  {https://arxiv.org/abs/1608.00754} {arXiv:1608.00754 [cond-mat.mtrl-sci]}
  \BibitemShut {NoStop}%
\bibitem [{\citenamefont {Azadi}\ \emph {et~al.}(2018)\citenamefont {Azadi},
  \citenamefont {Singh},\ and\ \citenamefont {Kühne}}]{Azadi2018}%
  \BibitemOpen
  \bibfield  {author} {\bibinfo {author} {\bibfnamefont {S.}~\bibnamefont
  {Azadi}}, \bibinfo {author} {\bibfnamefont {R.}~\bibnamefont {Singh}},\ and\
  \bibinfo {author} {\bibfnamefont {T.~D.}\ \bibnamefont {Kühne}},\ }\href
  {https://doi.org/10.1002/jcc.25104} {\bibfield  {journal} {\bibinfo
  {journal} {J. Comp. Chem.}\ }\textbf {\bibinfo {volume} {39}},\ \bibinfo
  {pages} {262} (\bibinfo {year} {2018})}\BibitemShut {NoStop}%
\bibitem [{\citenamefont {Norman}\ and\ \citenamefont
  {Saitov}(2019)}]{Norman2019}%
  \BibitemOpen
  \bibfield  {author} {\bibinfo {author} {\bibfnamefont {G.~E.}\ \bibnamefont
  {Norman}}\ and\ \bibinfo {author} {\bibfnamefont {I.~M.}\ \bibnamefont
  {Saitov}},\ }\href {https://doi.org/10.1134/S1028335819040116} {\bibfield
  {journal} {\bibinfo  {journal} {Dokl. Phys.}\ }\textbf {\bibinfo {volume}
  {64}},\ \bibinfo {pages} {145} (\bibinfo {year} {2019})}\BibitemShut
  {NoStop}%
\bibitem [{\citenamefont {Saitov}(2019)}]{Saitov2019}%
  \BibitemOpen
  \bibfield  {author} {\bibinfo {author} {\bibfnamefont {I.~M.}\ \bibnamefont
  {Saitov}},\ }\href {https://doi.org/10.1134/S0021364019150116} {\bibfield
  {journal} {\bibinfo  {journal} {J. Exp. Theor. Phys. Lett.}\ }\textbf
  {\bibinfo {volume} {110}},\ \bibinfo {pages} {206} (\bibinfo {year}
  {2019})}\BibitemShut {NoStop}%
\bibitem [{\citenamefont {Saitov}(2020)}]{Saitov2020}%
  \BibitemOpen
  \bibfield  {author} {\bibinfo {author} {\bibfnamefont {I.~M.}\ \bibnamefont
  {Saitov}},\ }\href {https://doi.org/10.1134/S1063776120010094} {\bibfield
  {journal} {\bibinfo  {journal} {J. Exp. Theor. Phys.}\ }\textbf {\bibinfo
  {volume} {130}},\ \bibinfo {pages} {423} (\bibinfo {year}
  {2020})}\BibitemShut {NoStop}%
\bibitem [{\citenamefont {Norman}\ and\ \citenamefont
  {Saitov}(2020)}]{Norman2020}%
  \BibitemOpen
  \bibfield  {author} {\bibinfo {author} {\bibfnamefont {G.~E.}\ \bibnamefont
  {Norman}}\ and\ \bibinfo {author} {\bibfnamefont {I.~M.}\ \bibnamefont
  {Saitov}},\ }\href {https://doi.org/10.1134/S0021364020030091} {\bibfield
  {journal} {\bibinfo  {journal} {J. Exp. Theor. Phys. Lett.}\ }\textbf
  {\bibinfo {volume} {111}},\ \bibinfo {pages} {162} (\bibinfo {year}
  {2020})}\BibitemShut {NoStop}%
\bibitem [{\citenamefont {Dorofeev}\ and\ \citenamefont
  {Fal’kovskii}(1984)}]{Dorofeev1984}%
  \BibitemOpen
  \bibfield  {author} {\bibinfo {author} {\bibfnamefont {E.~A.}\ \bibnamefont
  {Dorofeev}}\ and\ \bibinfo {author} {\bibfnamefont {L.~A.}\ \bibnamefont
  {Fal’kovskii}},\ }\href@noop {} {\bibfield  {journal} {\bibinfo  {journal}
  {Sov. J. Exp. Theor. Phys.}\ }\textbf {\bibinfo {volume} {60}},\ \bibinfo
  {pages} {1273} (\bibinfo {year} {1984})}\BibitemShut {NoStop}%
\bibitem [{\citenamefont {Brown}\ \emph {et~al.}(2015)\citenamefont {Brown},
  \citenamefont {Semeniuk}, \citenamefont {Vasiljkovic},\ and\ \citenamefont
  {Grosche}}]{Brown2015}%
  \BibitemOpen
  \bibfield  {author} {\bibinfo {author} {\bibfnamefont {P.}~\bibnamefont
  {Brown}}, \bibinfo {author} {\bibfnamefont {K.}~\bibnamefont {Semeniuk}},
  \bibinfo {author} {\bibfnamefont {A.}~\bibnamefont {Vasiljkovic}},\ and\
  \bibinfo {author} {\bibfnamefont {F.~M.}\ \bibnamefont {Grosche}},\ }\href
  {https://doi.org/10.1016/j.phpro.2015.12.005} {\bibfield  {journal} {\bibinfo
   {journal} {Phys. Procedia}\ }\textbf {\bibinfo {volume} {75}},\ \bibinfo
  {pages} {29} (\bibinfo {year} {2015})}\BibitemShut {NoStop}%
\bibitem [{\citenamefont {Shimizu}(2018)}]{Shimizu2018}%
  \BibitemOpen
  \bibfield  {author} {\bibinfo {author} {\bibfnamefont {K.}~\bibnamefont
  {Shimizu}},\ }\href {https://doi.org/10.1016/j.physc.2018.05.012} {\bibfield
  {journal} {\bibinfo  {journal} {Phys. C}\ }\textbf {\bibinfo {volume}
  {552}},\ \bibinfo {pages} {30} (\bibinfo {year} {2018})}\BibitemShut
  {NoStop}%
\bibitem [{\citenamefont {Eremets}\ \emph {et~al.}(2000)\citenamefont
  {Eremets}, \citenamefont {Gregoryanz}, \citenamefont {Struzhkin},
  \citenamefont {Mao}, \citenamefont {Hemley}, \citenamefont {Mulders},\ and\
  \citenamefont {Zimmerman}}]{Eremets2000}%
  \BibitemOpen
  \bibfield  {author} {\bibinfo {author} {\bibfnamefont {M.~I.}\ \bibnamefont
  {Eremets}}, \bibinfo {author} {\bibfnamefont {E.~A.}\ \bibnamefont
  {Gregoryanz}}, \bibinfo {author} {\bibfnamefont {V.~V.}\ \bibnamefont
  {Struzhkin}}, \bibinfo {author} {\bibfnamefont {H.-K.}\ \bibnamefont {Mao}},
  \bibinfo {author} {\bibfnamefont {R.~J.}\ \bibnamefont {Hemley}}, \bibinfo
  {author} {\bibfnamefont {N.}~\bibnamefont {Mulders}},\ and\ \bibinfo {author}
  {\bibfnamefont {N.~M.}\ \bibnamefont {Zimmerman}},\ }\href
  {https://doi.org/10.1103/PhysRevLett.85.2797} {\bibfield  {journal} {\bibinfo
   {journal} {Phys. Rev. Lett.}\ }\textbf {\bibinfo {volume} {85}},\ \bibinfo
  {pages} {2797} (\bibinfo {year} {2000})}\BibitemShut {NoStop}%
\bibitem [{\citenamefont {Koufos}\ and\ \citenamefont
  {Papaconstantopoulos}(2015)}]{Koufos2015}%
  \BibitemOpen
  \bibfield  {author} {\bibinfo {author} {\bibfnamefont {A.~P.}\ \bibnamefont
  {Koufos}}\ and\ \bibinfo {author} {\bibfnamefont {D.~A.}\ \bibnamefont
  {Papaconstantopoulos}},\ }\href@noop {} {\bibfield  {journal} {\bibinfo
  {journal} {J. Supercond. Novel Magn.}\ }\textbf {\bibinfo {volume} {28}},\
  \bibinfo {pages} {3525} (\bibinfo {year} {2015})}\BibitemShut {NoStop}%
\bibitem [{\citenamefont {Dogan}\ \emph {et~al.}(2021)\citenamefont {Dogan},
  \citenamefont {Oh},\ and\ \citenamefont {Cohen}}]{Dogan2021}%
  \BibitemOpen
  \bibfield  {author} {\bibinfo {author} {\bibfnamefont {M.}~\bibnamefont
  {Dogan}}, \bibinfo {author} {\bibfnamefont {S.}~\bibnamefont {Oh}},\ and\
  \bibinfo {author} {\bibfnamefont {M.~L.}\ \bibnamefont {Cohen}},\ }\href
  {https://doi.org/10.1088/1361-648X/abba8a} {\bibfield  {journal} {\bibinfo
  {journal} {J. Phys.: Condens. Matter}\ }\textbf {\bibinfo {volume} {33}},\
  \bibinfo {eid} {03LT01} (\bibinfo {year} {2021})},\ \Eprint
  {https://arxiv.org/abs/2006.16432} {arXiv:2006.16432 [cond-mat.mtrl-sci]}
  \BibitemShut {NoStop}%
\bibitem [{\citenamefont {Kresse}\ and\ \citenamefont
  {Hafner}(1993)}]{Kresse1993}%
  \BibitemOpen
  \bibfield  {author} {\bibinfo {author} {\bibfnamefont {G.}~\bibnamefont
  {Kresse}}\ and\ \bibinfo {author} {\bibfnamefont {J.}~\bibnamefont
  {Hafner}},\ }\href {https://doi.org/10.1103/PhysRevB.47.558} {\bibfield
  {journal} {\bibinfo  {journal} {Phys. Rev. B}\ }\textbf {\bibinfo {volume}
  {47}},\ \bibinfo {pages} {558} (\bibinfo {year} {1993})}\BibitemShut
  {NoStop}%
\bibitem [{\citenamefont {Kresse}\ and\ \citenamefont
  {Hafner}(1994)}]{Kresse1994}%
  \BibitemOpen
  \bibfield  {author} {\bibinfo {author} {\bibfnamefont {G.}~\bibnamefont
  {Kresse}}\ and\ \bibinfo {author} {\bibfnamefont {J.}~\bibnamefont
  {Hafner}},\ }\href {https://doi.org/10.1103/PhysRevB.49.14251} {\bibfield
  {journal} {\bibinfo  {journal} {Phys. Rev. B}\ }\textbf {\bibinfo {volume}
  {49}},\ \bibinfo {pages} {14251} (\bibinfo {year} {1994})}\BibitemShut
  {NoStop}%
\bibitem [{\citenamefont {Kresse}\ and\ \citenamefont
  {Furthm\"uller}(1996{\natexlab{a}})}]{Kresse1996}%
  \BibitemOpen
  \bibfield  {author} {\bibinfo {author} {\bibfnamefont {G.}~\bibnamefont
  {Kresse}}\ and\ \bibinfo {author} {\bibfnamefont {J.}~\bibnamefont
  {Furthm\"uller}},\ }\href {https://doi.org/10.1103/PhysRevB.54.11169}
  {\bibfield  {journal} {\bibinfo  {journal} {Phys. Rev. B}\ }\textbf {\bibinfo
  {volume} {54}},\ \bibinfo {pages} {11169} (\bibinfo {year}
  {1996}{\natexlab{a}})}\BibitemShut {NoStop}%
\bibitem [{\citenamefont {Kresse}\ and\ \citenamefont
  {Furthm\"uller}(1996{\natexlab{b}})}]{KresseFurthmueller1996}%
  \BibitemOpen
  \bibfield  {author} {\bibinfo {author} {\bibfnamefont {G.}~\bibnamefont
  {Kresse}}\ and\ \bibinfo {author} {\bibfnamefont {J.}~\bibnamefont
  {Furthm\"uller}},\ }\href {https://doi.org/10.1016/0927-0256(96)00008-0}
  {\bibfield  {journal} {\bibinfo  {journal} {Comp. Mat. Sci.}\ }\textbf
  {\bibinfo {volume} {6}},\ \bibinfo {pages} {15} (\bibinfo {year}
  {1996}{\natexlab{b}})}\BibitemShut {NoStop}%
\bibitem [{\citenamefont {{Nos{\'e}}}(1984)}]{Nose1984}%
  \BibitemOpen
  \bibfield  {author} {\bibinfo {author} {\bibfnamefont {S.}~\bibnamefont
  {{Nos{\'e}}}},\ }\href {https://doi.org/10.1063/1.447334} {\bibfield
  {journal} {\bibinfo  {journal} {J. Chem. Phys.}\ }\textbf {\bibinfo {volume}
  {81}},\ \bibinfo {pages} {511} (\bibinfo {year} {1984})}\BibitemShut
  {NoStop}%
\bibitem [{\citenamefont {Hoover}(1985)}]{Hoover1985}%
  \BibitemOpen
  \bibfield  {author} {\bibinfo {author} {\bibfnamefont {W.~G.}\ \bibnamefont
  {Hoover}},\ }\href {https://doi.org/10.1103/PhysRevA.31.1695} {\bibfield
  {journal} {\bibinfo  {journal} {Phys. Rev. A}\ }\textbf {\bibinfo {volume}
  {31}},\ \bibinfo {pages} {1695} (\bibinfo {year} {1985})}\BibitemShut
  {NoStop}%
\bibitem [{\citenamefont {Perdew}\ \emph {et~al.}(1996)\citenamefont {Perdew},
  \citenamefont {Burke},\ and\ \citenamefont {Ernzerhof}}]{Perdew1996}%
  \BibitemOpen
  \bibfield  {author} {\bibinfo {author} {\bibfnamefont {J.~P.}\ \bibnamefont
  {Perdew}}, \bibinfo {author} {\bibfnamefont {K.}~\bibnamefont {Burke}},\ and\
  \bibinfo {author} {\bibfnamefont {M.}~\bibnamefont {Ernzerhof}},\ }\href
  {https://doi.org/10.1103/PhysRevLett.77.3865} {\bibfield  {journal} {\bibinfo
   {journal} {Phys. Rev. Lett.}\ }\textbf {\bibinfo {volume} {77}},\ \bibinfo
  {pages} {3865} (\bibinfo {year} {1996})}\BibitemShut {NoStop}%
\bibitem [{\citenamefont {Stokes}\ and\ \citenamefont
  {Hatch}(2005)}]{Stokes2005}%
  \BibitemOpen
  \bibfield  {author} {\bibinfo {author} {\bibfnamefont {H.~T.}\ \bibnamefont
  {Stokes}}\ and\ \bibinfo {author} {\bibfnamefont {D.~M.}\ \bibnamefont
  {Hatch}},\ }\href@noop {} {\bibfield  {journal} {\bibinfo  {journal} {J.
  Appl. Crystallogr.}\ }\textbf {\bibinfo {volume} {38}},\ \bibinfo {pages}
  {237} (\bibinfo {year} {2005})}\BibitemShut {NoStop}%
\bibitem [{\citenamefont {Heyd}\ \emph {et~al.}(2003)\citenamefont {Heyd},
  \citenamefont {Scuseria},\ and\ \citenamefont {Ernzerhof}}]{Heyd2003}%
  \BibitemOpen
  \bibfield  {author} {\bibinfo {author} {\bibfnamefont {J.}~\bibnamefont
  {Heyd}}, \bibinfo {author} {\bibfnamefont {G.~E.}\ \bibnamefont {Scuseria}},\
  and\ \bibinfo {author} {\bibfnamefont {M.}~\bibnamefont {Ernzerhof}},\ }\href
  {https://doi.org/10.1063/1.1564060} {\bibfield  {journal} {\bibinfo
  {journal} {J. Chem. Phys.}\ }\textbf {\bibinfo {volume} {118}},\ \bibinfo
  {pages} {8207} (\bibinfo {year} {2003})}\BibitemShut {NoStop}%
\bibitem [{\citenamefont {Monkhorst}\ and\ \citenamefont
  {Pack}(1976)}]{Monkhorst1976}%
  \BibitemOpen
  \bibfield  {author} {\bibinfo {author} {\bibfnamefont {H.~J.}\ \bibnamefont
  {Monkhorst}}\ and\ \bibinfo {author} {\bibfnamefont {J.~D.}\ \bibnamefont
  {Pack}},\ }\href {https://doi.org/10.1103/PhysRevB.13.5188} {\bibfield
  {journal} {\bibinfo  {journal} {Phys. Rev. B}\ }\textbf {\bibinfo {volume}
  {13}},\ \bibinfo {pages} {5188} (\bibinfo {year} {1976})}\BibitemShut
  {NoStop}%
\bibitem [{\citenamefont {Ziman}(1972)}]{ziman_1972}%
  \BibitemOpen
  \bibfield  {author} {\bibinfo {author} {\bibfnamefont {J.~M.}\ \bibnamefont
  {Ziman}},\ }\href {https://doi.org/10.1017/CBO9781139644075} {\emph {\bibinfo
  {title} {Principles of the Theory of Solids}}},\ \bibinfo {edition} {2nd}\
  ed.\ (\bibinfo  {publisher} {Cambridge University Press},\ \bibinfo {year}
  {1972})\BibitemShut {NoStop}%
\bibitem [{\citenamefont {Ashcroft}\ \emph {et~al.}(1976)\citenamefont
  {Ashcroft}, \citenamefont {W}, \citenamefont {Ashcroft}, \citenamefont
  {Mermin}, \citenamefont {Mermin}, \citenamefont {Mermin},\ and\ \citenamefont
  {Company}}]{ashcroft1976solid}%
  \BibitemOpen
  \bibfield  {author} {\bibinfo {author} {\bibfnamefont {N.}~\bibnamefont
  {Ashcroft}}, \bibinfo {author} {\bibfnamefont {A.}~\bibnamefont {W}},
  \bibinfo {author} {\bibfnamefont {W.}~\bibnamefont {Ashcroft}}, \bibinfo
  {author} {\bibfnamefont {N.}~\bibnamefont {Mermin}}, \bibinfo {author}
  {\bibfnamefont {N.}~\bibnamefont {Mermin}}, \bibinfo {author} {\bibfnamefont
  {D.}~\bibnamefont {Mermin}},\ and\ \bibinfo {author} {\bibfnamefont {B.~P.}\
  \bibnamefont {Company}},\ }\href
  {https://books.google.nl/books?id=1C9HAQAAIAAJ} {\emph {\bibinfo {title}
  {Solid State Physics}}},\ HRW international editions\ (\bibinfo  {publisher}
  {Holt, Rinehart and Winston},\ \bibinfo {year} {1976})\BibitemShut {NoStop}%
\bibitem [{\citenamefont {Volkov}\ and\ \citenamefont
  {Fal’kovskii}(1983)}]{Volkov1983}%
  \BibitemOpen
  \bibfield  {author} {\bibinfo {author} {\bibfnamefont {B.~A.}\ \bibnamefont
  {Volkov}}\ and\ \bibinfo {author} {\bibfnamefont {L.~A.}\ \bibnamefont
  {Fal’kovskii}},\ }\href@noop {} {\bibfield  {journal} {\bibinfo  {journal}
  {Sov. J. Exp. Theor. Phys.}\ }\textbf {\bibinfo {volume} {58}},\ \bibinfo
  {pages} {1239} (\bibinfo {year} {1983})}\BibitemShut {NoStop}%
\bibitem [{\citenamefont {Fal’kovskii}(1986{\natexlab{a}})}]{Falkovskii1986}%
  \BibitemOpen
  \bibfield  {author} {\bibinfo {author} {\bibfnamefont {L.~A.}\ \bibnamefont
  {Fal’kovskii}},\ }\href@noop {} {\bibfield  {journal} {\bibinfo  {journal}
  {Sov. Phys. Usp.}\ }\textbf {\bibinfo {volume} {29}},\ \bibinfo {pages} {577}
  (\bibinfo {year} {1986}{\natexlab{a}})}\BibitemShut {NoStop}%
\bibitem [{\citenamefont
  {Fal’kovskii}(1986{\natexlab{b}})}]{Falkovskii1986a}%
  \BibitemOpen
  \bibfield  {author} {\bibinfo {author} {\bibfnamefont {L.}~\bibnamefont
  {Fal’kovskii}},\ }\href@noop {} {\bibfield  {journal} {\bibinfo  {journal}
  {Sov. Phys. Solid State}\ }\textbf {\bibinfo {volume} {28}},\ \bibinfo
  {pages} {146} (\bibinfo {year} {1986}{\natexlab{b}})}\BibitemShut {NoStop}%
\bibitem [{\citenamefont {Wan}\ \emph {et~al.}(2011)\citenamefont {Wan},
  \citenamefont {Turner}, \citenamefont {Vishwanath},\ and\ \citenamefont
  {Savrasov}}]{Wan2011}%
  \BibitemOpen
  \bibfield  {author} {\bibinfo {author} {\bibfnamefont {X.}~\bibnamefont
  {Wan}}, \bibinfo {author} {\bibfnamefont {A.~M.}\ \bibnamefont {Turner}},
  \bibinfo {author} {\bibfnamefont {A.}~\bibnamefont {Vishwanath}},\ and\
  \bibinfo {author} {\bibfnamefont {S.~Y.}\ \bibnamefont {Savrasov}},\ }\href
  {https://doi.org/10.1103/PhysRevB.83.205101} {\bibfield  {journal} {\bibinfo
  {journal} {Phys. Rev. B}\ }\textbf {\bibinfo {volume} {83}},\ \bibinfo {eid}
  {205101} (\bibinfo {year} {2011})},\ \Eprint
  {https://arxiv.org/abs/1007.0016} {arXiv:1007.0016 [cond-mat.str-el]}
  \BibitemShut {NoStop}%
\bibitem [{\citenamefont {Weng}\ \emph {et~al.}(2015)\citenamefont {Weng},
  \citenamefont {Fang}, \citenamefont {Fang}, \citenamefont {Bernevig},\ and\
  \citenamefont {Dai}}]{Weng2015}%
  \BibitemOpen
  \bibfield  {author} {\bibinfo {author} {\bibfnamefont {H.}~\bibnamefont
  {Weng}}, \bibinfo {author} {\bibfnamefont {C.}~\bibnamefont {Fang}}, \bibinfo
  {author} {\bibfnamefont {Z.}~\bibnamefont {Fang}}, \bibinfo {author}
  {\bibfnamefont {B.~A.}\ \bibnamefont {Bernevig}},\ and\ \bibinfo {author}
  {\bibfnamefont {X.}~\bibnamefont {Dai}},\ }\href
  {https://doi.org/10.1103/PhysRevX.5.011029} {\bibfield  {journal} {\bibinfo
  {journal} {Phys. Rev. X}\ }\textbf {\bibinfo {volume} {5}},\ \bibinfo {eid}
  {011029} (\bibinfo {year} {2015})},\ \Eprint
  {https://arxiv.org/abs/1501.00060} {arXiv:1501.00060 [cond-mat.mtrl-sci]}
  \BibitemShut {NoStop}%
\bibitem [{\citenamefont {Shekhar}\ \emph {et~al.}(2015)\citenamefont
  {Shekhar}, \citenamefont {Nayak}, \citenamefont {Sun}, \citenamefont
  {Schmidt}, \citenamefont {Nicklas}, \citenamefont {Leermakers}, \citenamefont
  {Zeitler}, \citenamefont {Skourski}, \citenamefont {Wosnitza}, \citenamefont
  {Liu}, \citenamefont {Chen}, \citenamefont {Schnelle}, \citenamefont
  {Borrmann}, \citenamefont {Grin}, \citenamefont {Felser},\ and\ \citenamefont
  {Yan}}]{Shekhar2015}%
  \BibitemOpen
  \bibfield  {author} {\bibinfo {author} {\bibfnamefont {C.}~\bibnamefont
  {Shekhar}}, \bibinfo {author} {\bibfnamefont {A.~K.}\ \bibnamefont {Nayak}},
  \bibinfo {author} {\bibfnamefont {Y.}~\bibnamefont {Sun}}, \bibinfo {author}
  {\bibfnamefont {M.}~\bibnamefont {Schmidt}}, \bibinfo {author} {\bibfnamefont
  {M.}~\bibnamefont {Nicklas}}, \bibinfo {author} {\bibfnamefont
  {I.}~\bibnamefont {Leermakers}}, \bibinfo {author} {\bibfnamefont
  {U.}~\bibnamefont {Zeitler}}, \bibinfo {author} {\bibfnamefont
  {Y.}~\bibnamefont {Skourski}}, \bibinfo {author} {\bibfnamefont
  {J.}~\bibnamefont {Wosnitza}}, \bibinfo {author} {\bibfnamefont
  {Z.}~\bibnamefont {Liu}}, \bibinfo {author} {\bibfnamefont {Y.}~\bibnamefont
  {Chen}}, \bibinfo {author} {\bibfnamefont {W.}~\bibnamefont {Schnelle}},
  \bibinfo {author} {\bibfnamefont {H.}~\bibnamefont {Borrmann}}, \bibinfo
  {author} {\bibfnamefont {Y.}~\bibnamefont {Grin}}, \bibinfo {author}
  {\bibfnamefont {C.}~\bibnamefont {Felser}},\ and\ \bibinfo {author}
  {\bibfnamefont {B.}~\bibnamefont {Yan}},\ }\href
  {https://doi.org/10.1038/nphys3372} {\bibfield  {journal} {\bibinfo
  {journal} {Nat. Phys.}\ }\textbf {\bibinfo {volume} {11}},\ \bibinfo {pages}
  {645} (\bibinfo {year} {2015})},\ \Eprint {https://arxiv.org/abs/1502.04361}
  {arXiv:1502.04361 [cond-mat.mtrl-sci]} \BibitemShut {NoStop}%
\end{thebibliography}

\end{document}